\documentclass[fleqn,usenatbib]{mnras}

\usepackage{newtxtext,newtxmath}

\usepackage[T1]{fontenc}

\DeclareRobustCommand{\VAN}[3]{#2}
\let\VANthebibliography\thebibliography
\def\thebibliography{\DeclareRobustCommand{\VAN}[3]{##3}\VANthebibliography}

\usepackage{xspace}
\usepackage{graphicx}	
\usepackage{amsmath}	

\newcommand{\ncode}[1]{{\sc #1}}
\newcommand{\sigame}{\ncode{s\'{i}game}\xspace}
\newcommand{\simba}{\ncode{simba}\xspace}
\newcommand{\gizmo}{\ncode{gizmo}\xspace}
\newcommand{\cloudy}{\ncode{cloudy}\xspace}

\title[Impact of scatter on \ion{C}{ii} power spectrum]{Revisiting the [\ion{C}{ii}]$_{\bmath{158\umu}\text{m}}$ line-intensity mapping power spectrum from the EoR using non-uniform line-luminosity scatter}

\author[C. S. Murmu and K. P. Olsen et al.]{Chandra Shekhar Murmu$^{1}$\thanks{E-mail: chandra0murmu@gmail.com (CSM)},
Karen P. Olsen$^{2}$,
Thomas R. Greve$^{3,4,5}$,
Suman Majumdar$^{1,6}$,
Kanan K. Datta$^{7}$,
\newauthor
Bryan R. Scott$^{8}$,
T. K. Daisy Leung$^{9}$,
Romeel Dav\'{e}$^{10,11,12}$,
Gerg\"{o} Popping$^{13}$,
Raul Ortega Ochoa$^{3,4}$,
\newauthor
David Vizgan$^{3,4,14}$, and
Desika Narayanan$^{15,16,3}$
\\
$^{1}$Department of Astronomy, Astrophysics and Space Engineering, Indian Institute of Technology Indore, Khandwa Rd., Simrol 453552, India\\
$^{2}$Department of Astronomy and Steward Observatory, University of Arizona, Tucson, AZ 85721, USA\\
$^{3}$Cosmic Dawn Center (DAWN)\\
$^{4}$National Space Institute, DTU Space, Technical University of Denmark, Elektrovej 327, DK-2800 Kgs.~Lyngby, Denmark\\
$^{5}$Department of Physics and Astronomy, University College London, Gower Street, London WC1E 6BT, UK\\
$^{6}$Department of Physics, Blackett Laboratory, Imperial College, London SW7 2AZ, UK\\
$^{7}$Relativity \& Cosmology Research Centre, Department of Physics, Jadavpur University, Kolkata 700032, India\\
$^{8}$Department of Physics \& Astronomy, University of California, Riverside, USA\\
$^{9}$Center for Computational Astrophysics, Flatiron Institute, 162 Fifth Avenue, New York, NY 10010, USA\\
$^{10}$Institute for Astronomy, Royal Observatory, Univeristy of Edinburgh, Edinburgh EH9 3HJ, UK\\
$^{11}$University of the Western Cape, Bellville, Cape Town 7535, South Africa\\
$^{12}$South African Astronomical Observatories, Observatory, Cape Town 7925, South Africa\\
$^{13}$European Southern Observatory, Karl-Schwarzschild-Stra{\ss}e, D-85748 Garching, Germany\\
$^{14}$Department of Astronomy, University of Illinois at Urbana-Champaign, 1002 West Green St., Urbana, IL 61801, USA\\
$^{15}$Department of Astronomy, University of Florida, 211 Bryant Space Sciences Center, Gainesville, FL 32611, USA\\
$^{16}$University of Florida Informatics Institute, 432 Newell Drive, CISE Bldg. E251, Gainesville, FL 32611, USA
}

\date{Accepted 2022 November 06. Received 2022 November 06; in original form 2021 December 04}

\pubyear{2021}

\begin{document}
\label{firstpage}
\pagerange{\pageref{firstpage}--\pageref{lastpage}}
\maketitle

\begin{abstract}
\noindent
Detecting the line-intensity mapping (LIM) signal from the galaxies of the Epoch of Reionization is an emerging tool to constrain their role in reionization. Ongoing and upcoming experiments target the signal fluctuations across the sky to reveal statistical and astrophysical properties of these galaxies via signal statistics, e.g., the power spectrum. Here, we revisit the [\ion{C}{ii}]$_{158 \umu\text{m}}$ LIM power spectrum under non-uniform line-luminosity scatter, which has a halo-mass variation of statistical properties. Line-luminosity scatter from a cosmological hydrodynamic and radiative transfer simulation of galaxies at $z=6$ is considered in this study. We test the robustness of different model frameworks that interpret the impact of the line-luminosity scatter on the signal statistics. We use a simple power-law model to fit the scatter and demonstrate that the mean luminosity-halo mass correlation fit cannot preserve the mean intensity of the LIM signal (hence the clustering power spectrum) under non-uniform scatter. In our case, the mean intensity changes by $\sim 48$ per cent compared to the mean correlation fit in contrast to the general case with semi-analytic scatter. However, we find that the prediction for the mean intensity from the most-probable fit can be modelled robustly, considering the generalized and more realistic non-uniform scatter. We also explore the possibility of diminishing luminosity bias under non-uniform scatter, affecting the clustering power spectrum, although this phenomenon might not be statistically significant. Therefore, we should adopt appropriate approaches that can consistently interpret the LIM power spectrum from observations.
\end{abstract}




\section{Introduction}\label{sec:intro}
\noindent
Probing the early galaxies from the Epoch of Reionization (EoR) is challenging, demanding very high resolution and sensitivities from the instruments trying to probe these galaxies. Observations with HST~\citep{Robertson+2015}, ALMA~\citep{Le_F_vre_2020}, SUBARU~\citep{Kashikawa+2006,Kashikawa+2011,Kashikawa+2014,Itoh+2018,Matsuoka+2018}, LAGER~\citep{Zheng+2017,Hu+2017,Hu+2019,Harish+2021,Wold+2022} have detected them in small numbers. Upcoming instruments like JWST~\citep{Steinhardt+2021} will further improve on the detections of these early galaxies. However, point-source detections through spectroscopy or photometry consume a significant amount of observational time and are thus expensive.

Line-intensity mapping (LIM) of galaxies \citep{Visbal_2010,Gong+2011a} is a possible solution for this, by which one can detect the integrated flux of atomic or molecular line emissions from numerous sources at once, without resolving them individually and with reduced sensitivity requirements. Moreover, it will significantly cut down the observational hours required to map large volumes of the sky and probe numerous galaxy samples, to infer about its properties. The line emission candidates promising for LIM experiments include [\ion{C}{ii}]$_{158 \umu\text{m}}$~\citep{Gong+2011b,Silva_2015,Dumitru_2019,Yue_2019,Sun_2021, Murmu+2021}, CO~\citep{Gong+2011a,Lidz+2011,Silva_2015,Li+2016,Breyesse+2017,Breyesse+2019,Ihle+2019,Moradinezhad_Dizgah+2019,Moradinezhad_Dizgah+2022}, Ly-$\alpha$~\citep{Visbal+2018,Heneka+2021} etc. Instruments like CONCERTO~\citep{lagache_2017,Concerto}, TIME~\citep{Crites_2014,Sun_2021}, FYST~\citep{Cothard_2020,CCAT-Prime_collaboration}, TIM~\citep{Vieira+2020} will be targeting the [\ion{C}{ii}]$_{158 \umu\text{m}}$ line. On the other hand, we have detections of the CO signal with COPSS~\citep{Keating+2015,Keating+2016}, COMAP~\citep{Ihle+2019,Chung+2021,Cleary+2021,Ihle+2021} and mmIME~\citep{Breysse+2022}. Future phases of the COMAP experiment will target the CO line to probe the EoR~\citep{Breysse+2021}. Similarly, SPHEREx~\citep{Visbal+2018,Heneka+2021,Cox+2022} and CDIM~\citep{Visbal+2018,Cooray+2019,Heneka+2021} will be mapping the universe with Ly-$\alpha$ detections. These experiments will be capturing the sky-fluctuations of the LIM signal, enabling us to estimate statistics, e.g., the power spectrum. It will aid us in understanding the large-scale distribution and the astrophysical properties of the ionizing sources from the EoR.

The presence of line-luminosity scatter will introduce a correction to the LIM power spectrum. Under line-luminosity scatter, we can model or interpret the change in the power spectrum in the following ways. One of the widely used ways is to interpret against the mean luminosity-halo mass ($\overline{L}(M_h, z)$) correlation function \citep{Li+2016,Schaan+2021,Yang+2022}. In this approach, the mean intensity of the LIM signal is preserved under line-luminosity scatter. Therefore, there is no change in the clustering (large-scale) power spectrum component, and only the shot-noise power is enhanced. The other approach \citep[followed by][]{Moradinezhad_Dizgah+2019,Moradinezhad_Dizgah+2022} is to use an $L(M_h,z)$ correlation fit, such that there is a change in the mean intensity and consequently in the clustering power. The deviation in the power spectrum at small $k$-modes under scatter can be modelled in terms of the scatter parameter $\mathbf{\sigma}$. We can utilize both models as long as we interpret the power spectrum accordingly. It can be done with scatter, modelled via a semi-analytical approach with a single scatter parameter uniform across halo-mass bins.

In this study, we have revisited various model-frameworks that can be used to interpret the impact of scatter on the 2-point statistic. We used line-luminosity scatter of the [\ion{C}{ii}]$_{158 \umu\text{m}}$ line emission, obtained from simulated data of cosmological hydrodynamic simulation \simba \citep[][]{Dave+2019,Leung+2020}. The outputs were post-processed with \cloudy \citep[][]{Ferland2013,Ferland2017} and \sigame \citep[][]{Olsen+2015,Olsen+2016,Olsen+2017,Leung+2020}. The scatter emerges naturally from the astrophysics implemented within this sophisticated simulation framework, and its statistical properties have halo-mass variation, making it non-uniform in nature. Our primary focus had been to explore whether all models can interpret the power spectrum under a more generalised and realistic line-luminosity scatter in a consistent and robust fashion. We demonstrate that a most-probable fit can robustly interpret the LIM signal's mean intensity and power spectrum.

We have organised this paper into the following sections: First, we discuss the astrophysical origins of the scatter in the [\ion{C}{ii}]$_{158 \umu\text{m}}$ luminosity in Section~\ref{sec:scatter}. The hydrodynamic simulation of the [\ion{C}{ii}]$_{158 \umu\text{m}}$ line emission is briefly discussed in Section~\ref{sec:sim} and fitting the scatter is described in Section~\ref{sec:fitting}. We then discuss the methods of remapping the line-luminosity scatter and estimating the LIM power spectrum in Section~\ref{sec:method}. The results are presented in Section~\ref{sec:results} and finally we summarize the paper in Section~\ref{sec:summary}. Throughout this work, we have adopted cosmological parameters $\Omega_{\text{m}}=0.3183\,,\Omega_\Lambda=0.6817\,,h=0.6704\,,\Omega_{\text{b}}h^2=0.022032\,,\sigma_8=0.8347\,,n_{\text{s}}=0.9619$, consistent with Planck+WP best-fit values \citep{Planck_XVI}.

\section[C II \texorpdfstring{158$\umu$m}{} line-luminosity scatter]{[C\,{\sevensize II}] \texorpdfstring{158$\bmath{\umu}$\lowercase{m}}{} line-luminosity scatter}\label{sec:scatter}
\noindent
The [\ion{C}{ii}]$_{158 \umu\text{m}}$ line-luminosity scatter originates from the collective dependence of $L_{\text{[CII]}}$ on various astrophysical factors such as star formation, metal enrichment, and different phases of the interstellar medium (ISM). The \sigame simulations by  \cite{Leung+2020} handle three ISM phases (ionized, atomic, and molecular), all of which emit [\ion{C}{ii}]$_{158 \umu\text{m}}$. The molecular phase, which makes up no more than $\sim 30$ per cent of the total ISM mass in the simulations, typically contributes by more than 50 per cent to the total [\ion{C}{ii}]$_{158 \umu\text{m}}$ emission, especially in massive galaxies \citep{Accurso2017, Vizgan2022}. Observationally, constraints on the contribution to the total [\ion{C}{ii}]$_{158 \umu\text{m}}$ emission from the molecular phase come from a survey of our Galaxy, which suggests that the combined dense PDR gas and CO-dark molecular gas make up $\sim 50\%$ of the total emission \citep{Pineda2014A&A...570}. Simulations also indicate that  the contribution to the [\ion{C}{ii}]$_{158 \umu\text{m}}$ emission from the ionized and atomic gas can be up to 50 per cent but decreases with increasing stellar mass and metallicity \citep{Accurso2017}. Although one would expect the [\ion{C}{ii}]$_{158 \umu\text{m}}$ emission to decrease at lower metallicities, this effect is negligible compared to the increase in CO photo-dissociation rate (and thus the available C$^+$ ions) that comes with lower metallicities \citep{Accurso2017}. The scatter in the $L_{\text{[CII]}}$ versus $M_{\text{halo}}$ correlation, therefore, primarily comes from the scatter in the relative mass distributions of the ISM phases, which are set by the specific star-formation rate and metallicity.
Although $L_{\text{[CII]}}$ is correlated to the host halo-mass, we would expect that, in reality, it isn't perfectly correlated. In the following subsections, we discuss the method of simulating the [\ion{C}{ii}]$_{158 \umu\text{m}}$ emission from galaxies and obtaining a one-to-one $L_{\text{[CII]}}$ versus $M_{\text{halo}}$ fit to the $L_{\text{[CII]}}$ scatter.

\subsection[Simulations of C II \texorpdfstring{158$\umu$m}{} emission]{Simulations of [C\,{\sevensize II}] \texorpdfstring{158$\bmath{\umu}$\lowercase{m}}{} emission}\label{sec:sim}
This work builds on the analysis of snapshots taken from the \simba suite of cosmological galaxy formation simulations, which themselves were evolved using the meshless finite mass hydrodynamics technique of \gizmo \citep{hopkins2015,hopkins2017,Dave+2019}. The \simba simulation set consists of three cubical volumes, 25, 50 and 100\,$h^{-1}$ cMpc ($h=0.678$) on each side, all of which are used in this work to search for galaxies at $z\sim6$. For each volume, a total of 1024$^3$ gas elements and 1024$^3$ dark matter particles are evolved from $z=249$. The galaxy properties in \simba have been compared to various observations across cosmic time \citep[][]{Thomas+2019,Appleby2020}, including the epoch of reionization \citep[][]{Wu+2020,Leung+2020}, and are in reasonable agreement. The sample of galaxies used here is the same as that presented in \cite{Leung+2020} and consists of 11,125 galaxies, with derived [\ion{C}{ii}]$_{158 \umu\text{m}}$ luminosities from 10$^{3.82}$ to 10$^{8.91}$ $L_\odot$. 

In order to derive the [\ion{C}{ii}]$_{158 \umu\text{m}}$ emission, the galaxy samples were post-processed with version 2 of the \sigame module \citep{Olsen+2017}.\footnote{\url{https://kpolsen.github.io/SIGAME/index.html}} This version of \sigame uses the spectral synthesis code \cloudy \citep[v17.01;][]{Ferland2013,Ferland2017} to model the line emission from the multi-phased ISM within each simulated galaxy. As input to \cloudy, \sigame uses physically motivated prescriptions to calculate the local interstellar radiation field (ISRF) spectrum, the cosmic ray (CR) ionization rate, and the gas density distribution of the ionized, atomic and molecular ISM phases (see \cite{Dave+2019} and \cite{Leung+2020} for  details on the \simba simulation and implementation of \sigame, respectively).

\subsection[Fitting the C II \texorpdfstring{158$\umu$m}{} line-luminosity scatter]{Fitting the [C\,{\sevensize II}] \texorpdfstring{158$\bmath{\umu}$\lowercase{m}}{} line-luminosity scatter}\label{sec:fitting}
\noindent
We derive the halo [\ion{C}{ii}]$_{158 \umu\text{m}}$ luminosity from the central galaxies' [\ion{C}{ii}]$_{158 \umu\text{m}}$ luminosity; these central galaxies are identified in the \simba simulation. We fit the [\ion{C}{ii}]$_{158 \umu\text{m}}$ line-luminosity scatter with a one-to-one $L_{\text{[CII]}}$-$M_{\text{halo}}$ model; this fit is used to interpret the impact of [\ion{C}{ii}]$_{158 \umu\text{m}}$ line-luminosity scatter on the power spectrum. The data from \simba simulation were used to obtain the fit.

We try to fit the scatter with multiple approaches. One of the ways we do it is by using all of the individual samples from the scatter data and applying least-squares minimization. The sample numbers were low enough ($\sim$10,000) to allow for an all-sample fit in this case, although it is not a versatile and reliable approach in general. Especially when dealing with large sample numbers, we need to resort to binned statistics.

\begin{figure*}
\centering
\includegraphics[width=0.8\textwidth]{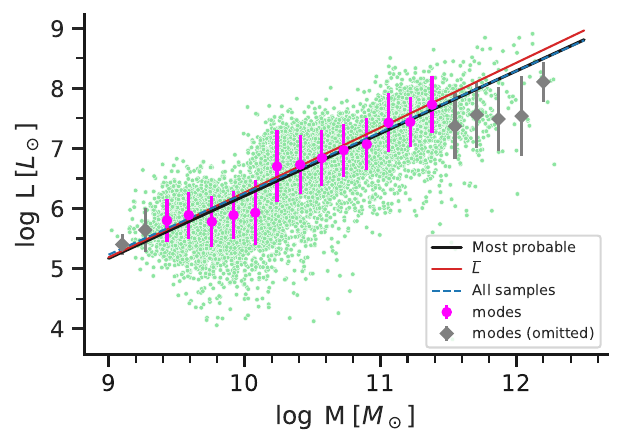}
\caption{$L_{\text{[CII]}}$ vs. $M_{\text{halo}}$ scatter from \simba simulation is shown here in \textit{green} dots, for $z=6$. The \textit{magenta} points are the most probable values or \textit{modes} of the $L_{\text{[CII]}}$ distribution in each halo-mass bin. The error bars are representative of the line-luminosity scatter in that halo-mass bin. The \textit{black solid, red solid and blue-dashed} lines represent the fits that are obtained in this work. The \textit{grey} points are excluded in the fitting analysis due to poor sample numbers.}
\label{fig:scatter}
\end{figure*}

\begin{figure*}
    \centering
    \includegraphics[width=\textwidth]{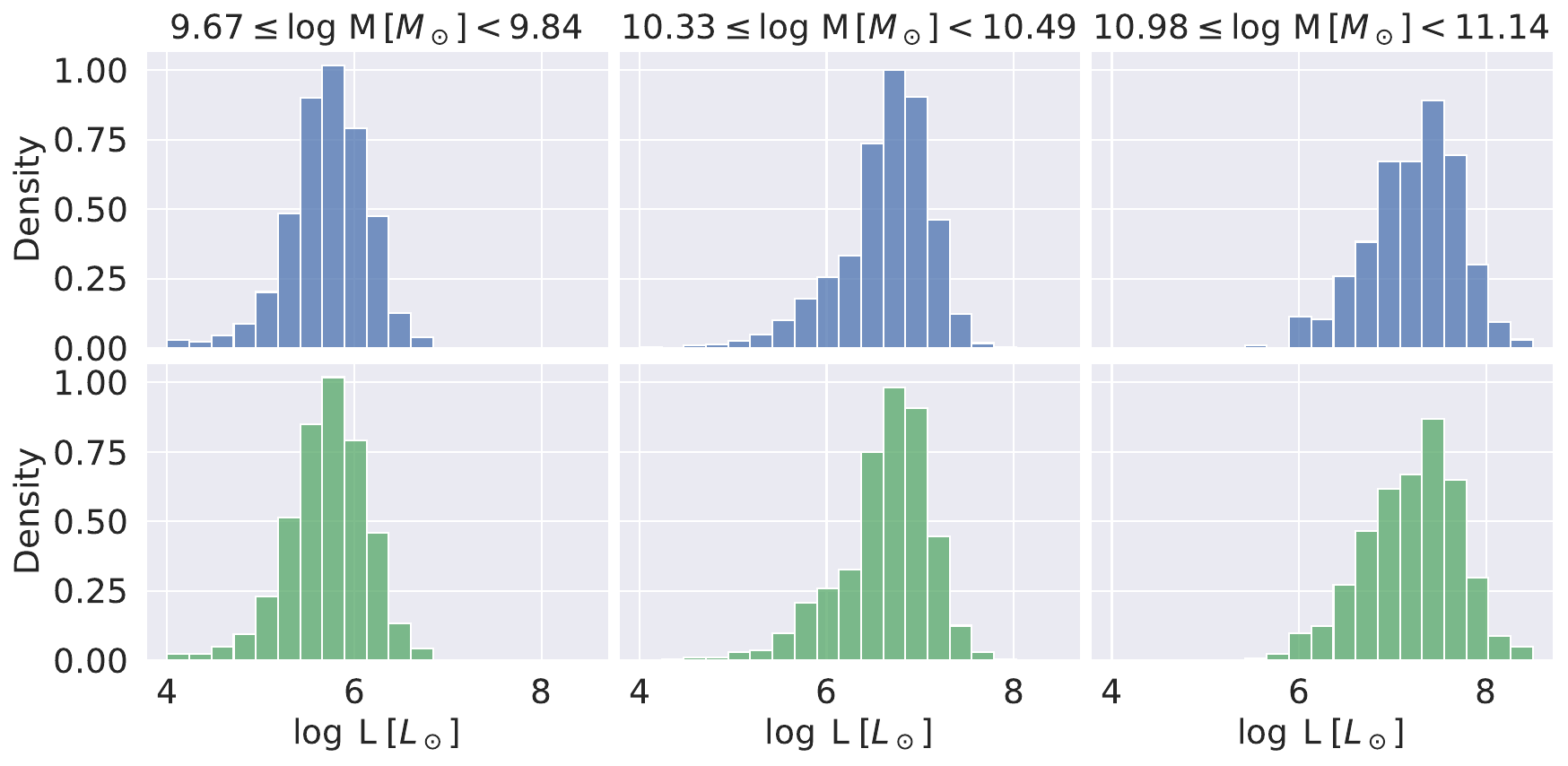}
    \caption{\textit{Top}: $L_{\text{[CII]}}$ distribution from \simba \textit{Bottom}: $L_{\text{[CII]}}$ distribution reproduced in N-body simulation from \simba data is shown.}
    \label{fig:bins}
\end{figure*}

The other way of fitting is to use binned data. We choose halo-mass bins at a logarithmic interval of $\sim 0.17$ dex. We estimated binned statistics such as the arithmetic mean and mode within each halo bin. To make the results more statistically significant, we ensure at least 200 samples from the lower and higher end of the halo-mass bins. The halo bins which do not have sufficient samples are dropped from the analysis (grey dots, Fig.~\ref{fig:scatter}). We try to apply least-squares minimization on the binned values (mean or modes) to obtain the correlation fits. Within the logarithmic halo-mass intervals, we estimated the arithmetic mean of $L_{\text{[CII]}}$, and with these bin values, we try to obtain a mean correlation fit. We used histograms of $\log L$ distribution (Fig.~\ref{fig:bins}, top panel) for each halo-mass bin of interest to evaluate the mode. From each halo-mass bin, we identified the peak of the histogram as our corresponding discretized mode, representing the most likely occurrence of $\log L$ in that bin. We used these modes in a least-square minimization procedure to obtain the corresponding most-probable fit.

Initially, we try to use the scaling relation from \cite{Leung+2020} to obtain the fits. This is given as
\begin{equation}
\log \bigg(\frac{L_{\text{[CII]}}}{L_\odot}\bigg) = C + a \log\bigg(\frac{M_{\text{h}}}{M_a}\bigg) + b \log\bigg(1+\frac{M_{\text{h}}}{M_b}\bigg),
\label{eq:Fit_Old}
\end{equation}
which corresponds to a double power law. The parameters for this model are, $C$, $a$, $M_a$, $b$ and $M_b$. However, given the simulation data that encompasses a particular halo-mass range ($10^9 \text{--} 10^{12}$), we do not obtain a robust fit using the relationship from \cite{Leung+2020}. We find that the fits are non-convergent, given an initial guess of parameters, and that the errors on the parameters are either unrealistically small or large. Instead, we use a simple power law model for our purpose, which has only two parameters. It is given as
\begin{equation}
\log \bigg(\frac{L_{\text{[CII]}}}{L_\odot}\bigg) = A + B\log\bigg(\frac{M_{\text{h}}}{M_\odot}\bigg),
\label{eq:Fit_New}
\end{equation}
with parameters $A$ and $B$. We do three fits with this model and tabulate the corresponding parameters in Table~\ref{tab:tab1}. The mean and the most probable fits are found to be convergent, given an initial guess for parameter values.

\begin{table}
\centering
\caption{Parameters and reduced chi-square for various fits: Most-probable, mean and all-sample are listed here.}
\label{tab:tab1}
\begin{tabular}{lccr}
\hline
Fits & A & B & $\chi^2/dof$ \\
\hline
    Most-probable & $-4.19\pm0.76$ & $1.04\pm0.07$ & $0.15$ \\
    Mean & $-4.53\pm0.61$ & $1.08\pm0.06$ & $0.08$ \\
    All-sample & $-3.95\pm0.07$ & $1.02\pm0.01$ & $1.14$\\
\hline
\end{tabular}
\end{table}

\begin{figure}
    \centering
    \includegraphics[width=\columnwidth]{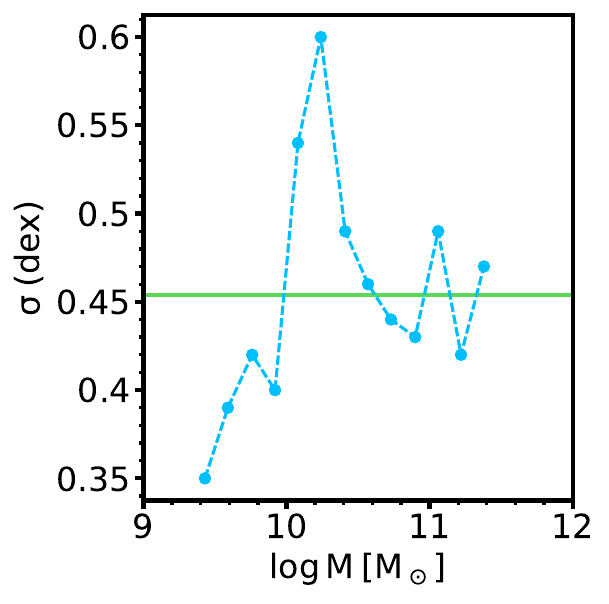}
    \caption{Fluctuations in scatter parameter ($\sigma$) for the [\ion{C}{ii}]$_{158\umu\text{m}}$ luminosity is shown across halo-mass bins. The horizontal \textit{green} line is the average value of the scatter parameter ($\sigma_{\rm avg})$.}
    \label{fig: sig_var}
\end{figure}

In Fig.~\ref{fig: sig_var}, we show the halo-mass variation of the scatter parameter. The \textit{green} line shows the average value of the parameter ($\sigma$). As mentioned earlier, the dataset from the \simba simulation consists of three different volumes, and consequently, the scatter parameter might have abrupt variations in the halo bins, where any two individual volumes overlap. Due to this, it is a bit difficult to ascertain the halo-mass versus scatter parameter trend. Nevertheless, within the individual volumes, we might find an overall variation. In the first couple of points, which lie within the smallest volume of the \simba simulation, we see the $\mathbf{\sigma}$ increasing. For the intermediate volume, this trend starts to fall off, and towards the end, we do not see any trend but fluctuations around the mean. Although this might not be the most accurate representation, we present a general scenario of how there might be a halo-mass-dependent variation of statistical properties of the scatter, as inferred from a hydrodynamic simulation. The following section describes the methods we used to estimate the [\ion{C}{ii}]$_{158 \umu\text{m}}$ LIM power spectrum.

\section[C II \texorpdfstring{158$\umu$m}{} LIM power spectrum with scatter]{[C\,{\sevensize II}] \texorpdfstring{158$\bmath{\umu}$\lowercase{m}}{} LIM power spectrum with scatter}\label{sec:method}
\noindent
The simulation volumes in \cite{Leung+2020} are not large enough to estimate the [\ion{C}{ii}]$_{158 \umu\text{m}}$ power spectrum with good statistical significance. Furthermore, the number of galaxies in individual volumes is not high enough as well. To deal with these problems, we remapped the scatter generated in the simulation suite of \cite{Leung+2020} in an N-body \citep{bharadwaj04} dark matter-only simulation with a volume of $215^3$ Mpc$^3$. The collapsed halos in the N-body simulation were identified with an FoF \citep{Mondal_2015} algorithm, with a halo-mass resolution of $\approx 10^9 M_\odot$.

The range of halo-mass in both the simulations are similar, which eased this exercise of reproducing the scatter. First, we divide the halo-mass range in into 20 log-arithmic bins. Then, within each halo-mass bin, we estimate the distribution of [\ion{C}{ii}]$_{158 \umu\text{m}}$ luminosity from the original scatter. The next step is to generate values of [\ion{C}{ii}]$_{158 \umu\text{m}}$ luminosity following the same distribution as the original one, within each halo-mass bin. One way of doing that could be to generate random $L_{\text{[CII]}}$ values using \textit{inverse transform sampling}, in which values are sampled from the empirical cumulative distribution function \citep{astroML,astroMLText}. However, here we modeled the original distribution from \simba with a \textit{piecewise constant distribution}. Given the information about the $\log L_{\text{[CII]}}$ histogram, such as boundaries of the bin intervals and the probability densities, within each halo-mass bin, we generate a random $\log L_{\text{[CII]}}$ value according to a probability density function given by

\begin{equation}
    P(\log L_{\text{[CII]}}|l_0,...,l_n,w_0,...,w_{n-1}) = \frac{w_i}{\sum_{k=0}^{n-1} w_k (l_{k+1} - l_k)} \, .
    \label{eq:Hist_Reprod}
\end{equation}

Here $n+1$ is the number of boundaries separating the intervals. $l_i$ are the boundaries of the intervals, with $0 \leq i \leq n$, and $w_i$ are the corresponding weights or the probability densities, with $0 \leq i < n$. For a given $L_{\text{[CII]}}$, it will satisfy $l_i \leq \log L_{\text{[CII]}} < l_{i+1}$, with $0 \leq i < n$. Thus, within a certain bin, uniform $\log L_{\text{[CII]}}$ values are generated with a certain associated weight, such that the overall distribution is reproduced in an almost exact fashion. This implementation is adopted here from the C++ standard library.

The histograms of the reproduced scatter are shown in green in the bottom panels of Fig. \ref{fig:bins}, corresponding to the method of equation~\ref{eq:Hist_Reprod}. We can see subtle differences between the original histogram (top panels of Fig. \ref{fig:bins}) and the reproduced one, although they are mostly similar.

\begin{figure}
    \centering
    \includegraphics[width=\columnwidth]{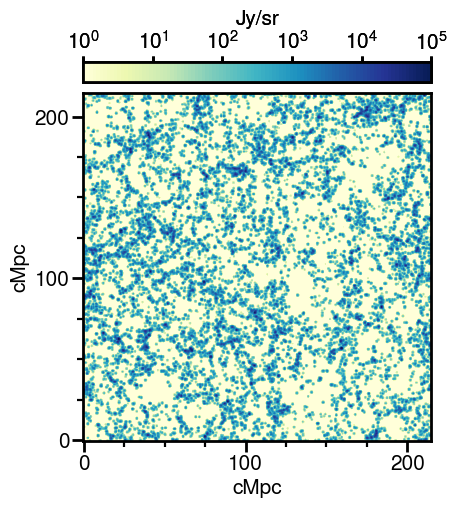}
    \caption{A snapshot of the [\ion{C}{ii}]$_{158 \umu\text{m}}$ intensity map with line-luminosity scatter at $z=6.$}
    \label{fig:maps}
\end{figure}

Using this method, we reproduced the actual scatter distribution present in the original simulation by \cite{Leung+2020}, in our N-body simulation. We also generated 1000 different realizations of such scatter distributions, and eventually that many number of [\ion{C}{ii}]$_{158 \umu\text{m}}$ intensity maps were also developed. Fig.~\ref{fig:maps} represents a snapshot of one such realization map.
Finally, the power spectra were computed for each of these intensity maps and averaged over, for the case with $L_{\text{[CII]}}$ scatter. The spatial distribution of the [\ion{C}{ii}]$_{158 \umu\text{m}}$ line emitters in our simulation volume is directly sampled to estimate the power spectrum.

\section{Results}\label{sec:results}
\subsection[Intensity of the C II \texorpdfstring{158$\umu$m}{} LIM signal]{Intensity of the [C\,{\sevensize II}] \texorpdfstring{158$\bmath{\umu}$\lowercase{m}}{} LIM signal}\label{sec:intensity}
We estimate the mean intensity of the LIM signal with and without scatter. The no-scatter LIM maps are generated by painting the halos with correlation functions obtained from different fitting approaches. We tabulate the results in Table~\ref{tab:tab2}.

\begin{table}
    \centering
    \caption{Mean intensities for each of the following specific cases at $z=6$ is tabulated.}
    \label{tab:tab2}
    \begin{tabular}{lr}
        \hline
        Case & Mean Intensity ($10^2$ Jy/sr)\\
        \hline
         Scatter &  $5.25$\\
         Most probable fit & $3.07$\\
         Mean fit & $3.55$\\
         All sample fit & $3.45$\\
         \hline
    \end{tabular}
\end{table}

There is an enhancement of $1.7$ in the mean intensity of the LIM signal with scatter compared to the most probable fit. For a given halo-mass bin, the average luminosity with log-scatter is

\begin{equation}
    \overline{L} = \int_{-\infty}^{\infty}d(\log L)L \times \frac{1}{\sqrt{2\pi}\sigma}\exp{-\frac{(\log L - \log \hat{L})^2}{2\sigma^2}},
    \label{eq:L_Avg1}
\end{equation}
which we can rewrite as

\begin{equation}
    \frac{\overline{L}(\sigma)}{\overline{L}_{\sigma=0}} = \int_{-\infty}^{\infty}dx \frac{10^{\sqrt{2}\sigma x}}{\sqrt{\pi}} \exp{-x^2} = 10^{\sigma^2\ln(10)/2}
     \label{eq:Ratio}
\end{equation}
\citep[see][]{Moradinezhad_Dizgah+2019}, with $\overline{L}_{\sigma=0}=\hat{L}$ being the average of $L$ for $\sigma=0$ (no scatter), and $\hat{L}$ is the average of $L$ in log-space. We find that the average $\sigma$ for the halo-mass bins considered in the fitting, from the \simba + \sigame results, is $0.45$ dex which, when put in equation~(\ref{eq:Ratio}), yields an approximate factor of enhancement in the mean intensity of $\sim 1.7$, in agreement with our simulation result. Therefore, using this simple model, we can interpret the mean intensity of the [\ion{C}{ii}]$_{158 \umu\text{m}}$ maps with log-normal line-luminosity scatter.

If the scatter were implemented with a semi-analytic model, we might have expected the mean intensity to remain preserved compared to this mean fit (see Appendix~\ref{sec: semi_analytic}). However, it is not straightforward in the presence of non-uniform scatter. In our case, the mean intensity changes by $\sim 48$ per cent compared with the mean correlation fit. This change is hard to interpret and model. We obtain a change of $\sim$ 52 per cent in the mean intensity with the all-sample fit. However, the drawback of using this fit is that it lacks statistical interpretation, unlike the others. Moreover, as mentioned earlier, this fit is not expected to be versatile (e.g., for large sample numbers) and reliable, especially since it lacks an underlying statistical model.

\subsection[C II \texorpdfstring{158$\umu$m}{} power spectrum]{[C\,{\sevensize II}] \texorpdfstring{158$\bmath{\umu}$\lowercase{m}}{} power spectrum}\label{sec:power_spectrum}

\begin{figure}
    \centering
    \includegraphics[width=\columnwidth]{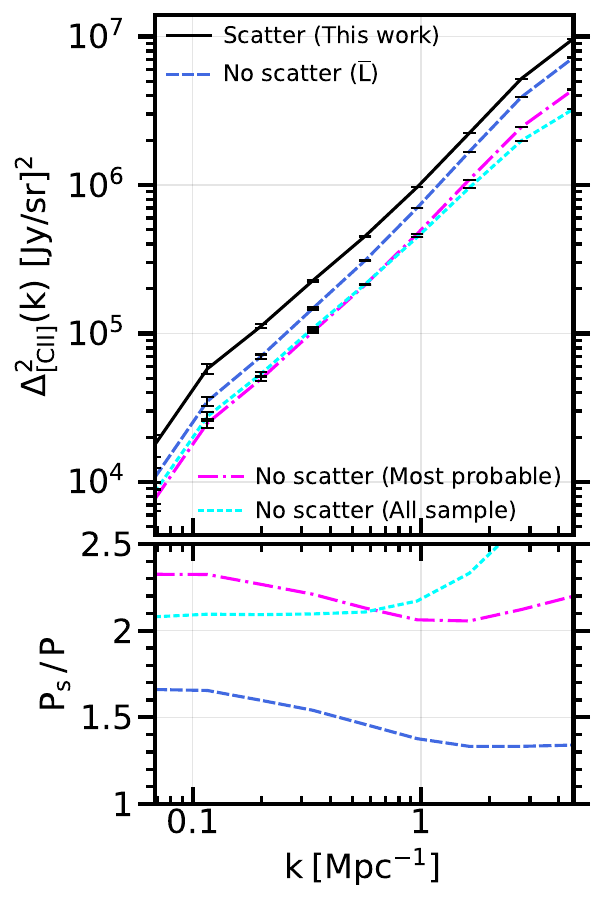}
    \caption{\textit{Top}: [\ion{C}{ii}]$_{158 \umu\text{m}}$ power spectrum at $z=6$ with and without $L_{\text{[CII]}}$ scatter. \textit{Bottom}: The ratio of the power spectrum with scatter compared to the no-scatter case is shown.}
    \label{fig:CII_power}
\end{figure}

We show the dimensionless power spectrum $\Delta^2(k) = k^3P(k)/2\pi^2$ in Fig.~\ref{fig:CII_power}, with $P(k)$ given as $\langle \tilde{\delta}^*(\mathbf{k^\prime})\tilde{
\delta}(\mathbf{k})\rangle = V\delta_{\mathbf{k^\prime},\mathbf{k}}P(k)$, where $\tilde{\delta}(\mathbf{k})$\footnote{This convention assumes a  definition of Fourier transform as\\ $\delta(\mathbf{x}) = \int \frac{d^3k}{(2\pi)^3}\exp{(i\mathbf{k}\cdot\mathbf{x})} \Tilde{\delta} (\mathbf{k})$} represent the signal fluctuations in Fourier-domain. $V$ is the volume of the box considered in the power-spectrum estimation. The large-scale power spectrum with line-luminosity scatter is enhanced by a factor of 2.3 -- 2.1, compared to the most probable fit. However, since the clustering power spectrum is expected to go as
\begin{equation}
    P^{\text{clus}}_{[\text{CII}]}(k, z) \propto \bar{I}^2_{[\text{CII}]}(z)\bar{b}^2_{[\text{CII}]}(z),
    \label{eq:Power_spec}
\end{equation}
with $\overline{I}^2_{\text{[CII]}}$ and $\overline{b}^2_{\text{[CII]}}$ being the mean [\ion{C}{ii}]$_{158 \umu\text{m}}$ intensity and bias, we should have expected an enhancement close to $1.7^2 \approx 2.9$. We try to reconcile this slight mismatch in Section~\ref{sec:bias}.

\begin{figure}
    \centering
    \includegraphics[width=\columnwidth]{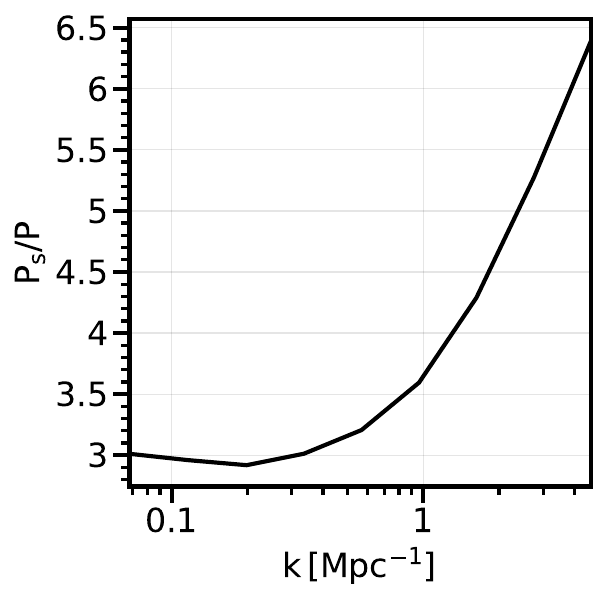}
    \caption{The impact of semi-analytic scatter on the power spectrum using model~(\ref{eq:Fit_New}) and parameters from the most-probable fit is shown here.}
    \label{fig:Semi-analytic_impact}
\end{figure}

We see that the impact of the scatter on the power spectrum (Fig.~\ref{fig:CII_power}) for the mean and most-probable fits, varies in a fashion which is different from what would be the case for scatter implemented with a semi-analytic approach. For reference, we show the impact of scatter when implemented via a semi-analytic method using equation~(\ref{eq: semi_analytic1}) in Fig.~\ref{fig:Semi-analytic_impact}. We use the model from equation~(\ref{eq:Fit_New}) with a $\sigma = 0.45$. Within the range of k-modes explored (0.1 -- 4 Mpc$^{-1}$), we see that the impact is initially almost constant but starts to rise steeply afterwards. The enhancement in the clustering power spectrum at low k-modes is less than the shot-noise power at large k-modes. It can be modelled with
\begin{equation}
f_{n,\sigma} = \int_{-\infty}^{\infty} dx \frac{10^{\sqrt{2}n\sigma x}}{\sqrt{\pi}}\exp-x^2 = 10^{n^2\sigma^2\ln(10)/2}
\end{equation}
\citep{Moradinezhad_Dizgah+2019}. $f_{1,\sigma}^2$ represents the enhancement in clustering power spectrum, where as $f_{2,\sigma}$ is the enhancement in the shot-noise component. However, we do not obtain this well-behaved enhancement for non-uniform scatter within the $k$-mode range we explore, which has varying scatter strength for halo-mass bins, in contrast to the semi-analytic case. The impact of shot-noise might dominate beyond the k-range probed here for these cases. The impact on the power spectrum behaves similarly to the semi-analytic case for the all-sample fit. However, the fit is not expected to be versatile and reliable.

\subsection{Luminosity-weighted halo bias}\label{sec:bias}
A plausible reason for the slight mismatch in power spectrum enhancement is the slight diminishing of the luminosity bias under non-uniform line-luminosity scatter. From equation~(\ref{eq:Power_spec}), we can follow that this decrement will suppress the maximum enhancement in the clustering power. The linear [\ion{C}{ii}]$_{158 \umu\text{m}}$ luminosity bias can be written as
\begin{equation}
     \bar{b}_{[\text{CII}]}(z) = \frac{\int dM \frac{dn}{dM} L_{[\text{CII}]}(M,z) b(M,z)}{\int dM \frac{dn}{dM} L_{[\text{CII}]}(M,z)},
\end{equation}
with $b(M,z)$ being the halo bias. When there is a non-uniform scatter, the binned average luminosity has stochastic fluctuations across halo-mass bins and does not follow any correlation function tightly (see Appendix~\ref{sec: semi_analytic}). It can thus introduce a overall decorrelation between the luminosity and the halo bias across the halo-mass bins, and therefore diminish the luminosity weighted halo bias slightly.

\begin{figure}
\centering
\includegraphics[width=\columnwidth]{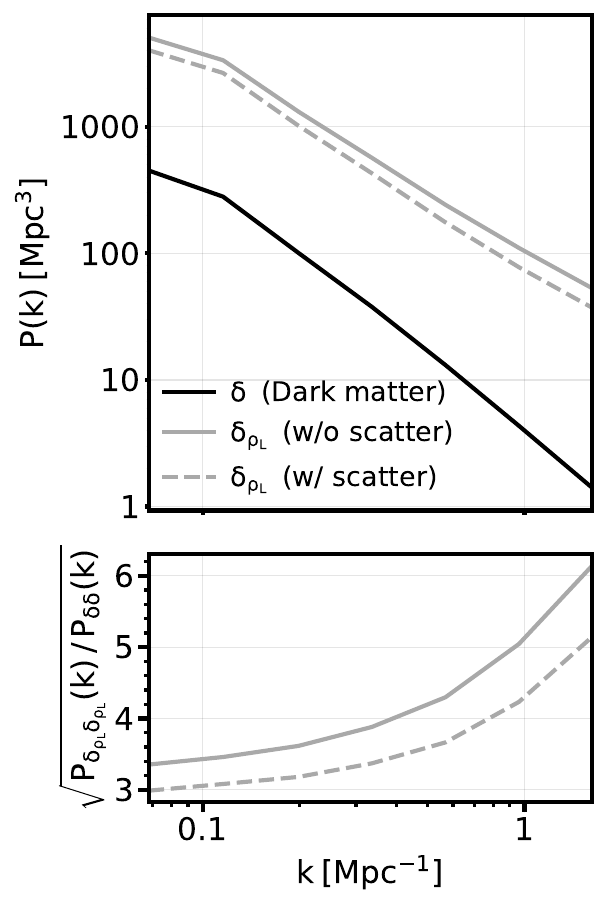}
\caption{\textit{Top panel}: Dark matter and luminosity fluctuations are shown for $z=6$. \textit{Bottom panel}: The estimated luminosity bias is shown.}
\label{fig: bias}
\end{figure}

We estimated the luminosity bias using the relation
\begin{equation}
    P_{\delta_{\rho_L}\delta_{\rho_L}} (k,z) = b^2_{\rm L}(k,z) P_{\delta\delta}(k,z),
    \label{eq:Delta_Fluc}
\end{equation}
to test this. Here, the $\delta_{\rho_L}$ represents fluctuations in the luminosity density defined as $\rho_L = \overline{\rho}_L(1+\delta_{\rho_L})$. $P_{\delta_{\rho_L}\delta_{\rho_L}}$ and $P_{\delta\delta}$ represents the power spectrum for the fluctuations corresponding to luminosity and dark matter. $b_{\rm L}$ is the luminosity bias. Fig.~\ref{fig: bias} shows the power spectrum for dark matter and luminosity fluctuations, corresponding to the scatter and no-scatter case. The corresponding bias is shown in the bottom panel, and we see a change of around $\sim$ 16 -- 11 per cent. Although this is not a very large change or statistically significant either, it remains as a possibility. Further studies are required to confirm this effect. When we try to reconcile this with the power spectrum, it matches the enhancement factor within 2.3 -- 2.1.

\section{Summary}\label{sec:summary}

We revisit the impact of line-luminosity scatter on the [\ion{C}{ii}]$_{158 \umu\text{m}}$ LIM power spectrum in this study. Line-luminosity scatter from a hydrodynamic simulation is considered, which differs from a simple semi-analytic one in its non-uniform nature (variation of statistical properties across halo-mass bins). Under this scenario, we test the robustness of various correlation fits that can meaningfully interpret the impact of line-luminosity scatter. We use a simple power law model to fit the [\ion{C}{ii}]$_{158 \umu\text{m}}$ line-luminosity scatter. The mean intensity for all-sample fit differs from the scatter case by 52 per cent. However, the fit obtained is not expected to be versatile and reliable since it lacks a meaningful, statistical interpretation. We find that the mean correlation fit produces mean intensity, which deviates from the scatter one by a large margin (48 per cent), although this shouldn't be the case if one deals with semi-analytic scatter (see Appendix~\ref{sec: semi_analytic}). Therefore, modelling the power spectrum with this fit becomes unreliable under non-uniform scatter. However, we find that the approach presented by \cite{Moradinezhad_Dizgah+2019} (recognized here as the most probable fit) provides a robust interpretation of the mean intensity (equation~(\ref{eq:Ratio})) and power spectrum, even under the generalized and realistic non-uniform line-luminosity scatter. This work demonstrates one example that the most probable fit might be the most reliable way to interpret the impact of line-luminosity scatter on the LIM signal statistics (e.g., mean intensity and power spectrum) compared to the other correlation fits considered.

The correlation fits we used in the analysis are obtained from a limited number of galaxy samples ($\sim$10,000) in our simulation. A higher sample number is desirable to get better statistical significance on the fits and thus the robustness of the conclusions presented here. Although this is generally difficult to achieve due to computational costs, increasing applications of emulators might become more relevant for these problems. Accurate reproduction of physics by emulators trained on hydro-simulations is thus the way forward to tackle these challenges. In this study, we limited our analysis to only $z=6$. There are interesting aspects that can be explored further, such as the variation of the non-uniformity in scatter with redshift. We do not separately model the duty cycle in our analysis. The inclusion of this parameter will change the scaling of the power spectrum, and the general conclusions presented in this work will remain unchanged. This concept is more relevant in studying time-based variation in the scatter. We hope to address these issues in future works.

From the observational perspective of the LIM signal, future experiments might be able to constrain some aspects (e.g. the slope) of the $L_{\text{[CII]}}$ - SFR and SFR - $M_{\text{halo}}$ relation \citep{Karoumpis+2021} from the power spectrum. However, the power spectrum alone might not be able to constrain the scatter information ($\sigma$). As we can understand from equation~(\ref{eq:Ratio}), the same power spectrum can result from different combinations of $L_{\text{[CII]}}$-$M_{\text{halo}}$ relation (whose amplitude can be parametrized by the log-space average of log-normal scatter or $\hat{L}$) and $\sigma$. Therefore, to better constrain the astrophysics of the LIM signal, one might need to use higher-order statistics, such as the bispectrum.

Although we primarily focus on the [\ion{C}{ii}]$_{158 \umu\text{m}}$ line emission, a similar analysis applies to the other relevant line emissions such as the CO, Ly-$\mathbf{\alpha}$ and [\ion{O}{iii}]$_{88 \umu\text{m}}$. The non-uniformity of scatter and its redshift evolution would be an interesting feature to study for these line emissions. We hope to explore this also in future works.

\section*{Acknowledgements}
We thank the referee for reviewing this article and providing vital feedback, which substantially improved the paper. CSM acknowledges funding from the Council of Scientific and Industrial Research (CSIR) via a CSIR-JRF fellowship, under the grant~09/1022(0080)/2019-EMR-I. KPO is funded by the National Aeronautics and Space Administration (NASA), under award No.~80NSSC19K1651. SM acknowledges financial support through the project titled ``Observing the Cosmic Dawn in Multicolour using Next Generation Telescopes'' funded by the Science and Engineering Research Board (SERB), Department of Science and Technology, Government of India through the Core Research Grant No. CRG/2021/004025. KKD acknowledges financial support from Science and Engineering Research Board (SERB), Government of India, through a MATRICS project grant, File No. MTR/2021/000384.
DN acknowledges the National Science Foundation (NSF) via AST-1909153 and NASA under award No.~80NSSC19K1651. The Cosmic Dawn Center of Excellence is funded by the Danish National Research Foundation under grant No.~140. Part of this research was done using the computing resources available to the Cosmology with Statistical Inference (CSI) research group at IIT Indore.

This research made use of arXiv (\url{https://arxiv.org}) research-sharing platfrom, NASA Astrophysics Data System Bibliographic Services and Crossref (\url{https://www.crossref.org}) citation services. The following softwares have been used: \texttt{N-body}\footnote{\url{https://github.com/rajeshmondal18/N-body}} \citep{bharadwaj04}, \texttt{FoF-Halo-Finder}\footnote{\url{https://github.com/rajeshmondal18/FoF-Halo-finder}} \citep{Mondal_2015}, \simba\footnote{\url{http://simba.roe.ac.uk/}} \citep{Dave+2019,Leung+2020}, \cloudy \citep{Ferland2013,Ferland2017}, \sigame\footnote{\url{https://github.com/kpolsen/SIGAME_v2}} \citep{Olsen+2015,Olsen+2016,Olsen+2017,Leung+2020}, \texttt{NumPy} \citep{harris2020array}, \texttt{pandas} \citep{reback2020pandas,mckinney-proc-scipy-2010}, \texttt{seaborn}\footnote{\url{https://seaborn.pydata.org/}} \citep{Waskom2021}, \texttt{ProPlot}\footnote{\url{https://proplot.readthedocs.io/en/stable/}}
\citep{luke_l_b_davis_2021_5495979}. The code used here to compute power spectrum is customized and adapted from \texttt{N-body} \citep{bharadwaj04} and \texttt{ReionYuga}\footnote{\url{https://github.com/rajeshmondal18/ReionYuga}} \citep{Choudhury_2009,Majumdar_2014,Mondal+2017}
\section*{Data Availability}
The simulated data underlying this work will be shared upon reasonable request to the corresponding author(s).


\bibliographystyle{mnras}
\bibliography{references} 




\appendix

\section{Implementation of semi-analytic scatter}
\label{sec: semi_analytic}
\begin{figure}
    \centering
    \includegraphics[width=\columnwidth]{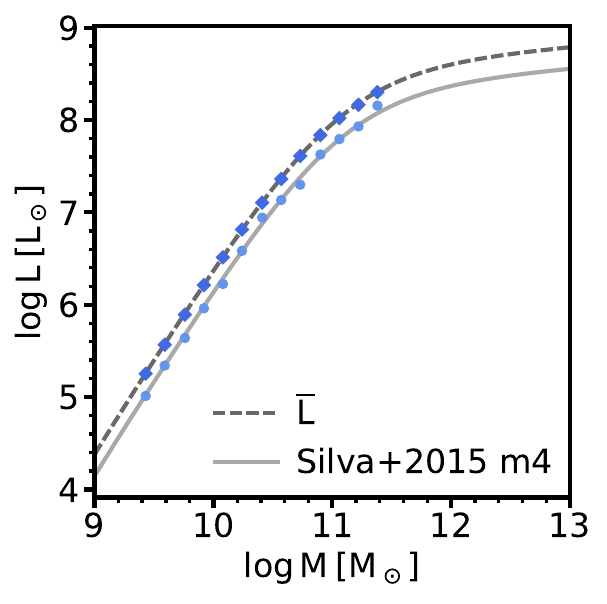}
    \caption{The solid grey line represents the m4 model \citep{Silva_2015} as the most probable fit. The corresponding mean correlation fit is shown in the dashed grey line. The mean fit is obtained by shifting the most probable fit using equation~(\ref{eq: semi_analytic2}). The corresponding points are the binned values (modes and arithmetic mean) for the correlation fits derived from the implemented line-luminosity scatter. We note that the points tightly follow the respective correlation fits.}
    \label{fig: semi_analytic}
\end{figure}

Here, we discuss how the semi-analytic scatter compares to a scatter obtained from hydrodynamic simulation. If we assume $L_{\rm m}(M,z)$ to be any correlation function between halo-mass and luminosity (corresponding to mode), then the log-scatter can be reproduced by
\begin{equation}
    \log L = \log L_{\rm m} + \mathcal{N}(0, \sigma^2),
    \label{eq: semi_analytic1}
\end{equation}
where $\mathcal{N}(0, \sigma^2)$ represents normal distribution with zero mean and $\sigma^2$ variance. The corresponding mean correlation function is then related to $L_{\rm m}$ by
\begin{equation}
    \log \overline{L} = \log L_{\rm m} + \frac{1}{2}\sigma^2\ln (10)
    \label{eq: semi_analytic2}
\end{equation}
\citep{Sun+2019,Yang+2022}. In this semi-analytic approach, since we use $L_{\rm m}$ as a base for generating scatter, the binned values of modes will lie very close to the correlation function. We illustrate this in Fig.~\ref{fig: semi_analytic}, using the model from Silva+2015 m4 (light-grey curve in Fig.~\ref{fig: semi_analytic}) as the $L_{\rm m}$. We use eq. ~\ref{eq: semi_analytic1} to generate scatter and estimate the binned values of modes (light-blue points in Fig.~\ref{fig: semi_analytic}), and we see that the correlation function almost passes through all the mode values. The same is true for the mean correlation function as well. So when we introduce scatter in this fashion, the binned values will tightly follow the corresponding correlation function.

However, the scatter drawn from the hydrodynamic simulation described is different. The origin of this is the more accurate astrophysics implemented in the simulation. The result is that the binned values of modes (or the mean, for that matter) do not follow the correlation function tightly. They fluctuate stochastically, which makes some aspects different. The relation~\ref{eq: semi_analytic2} does not hold necessarily, i.e., if $\sigma_{\rm avg}$ is the average value of the scatter parameter, then it is possible that
\begin{equation}
    \log \overline{L} \neq \log L_{\rm m} + \frac{1}{2}\sigma_{\rm avg}^2\ln (10).
\end{equation}
If we shift $\log L_{\rm m}$ by $\sigma_{{\rm avg}}^2\ln(10)/2$, it no longer corresponds to the mean correlation function. We demonstrate this by showing that an independently obtained mean correlation function doesn't keep the mean intensity invariant. In this perspective, the most-probable fit is a more robust approach to interpret the mean intensity and power spectrum of the LIM signal, even under non-uniform line-luminosity scatter.

\bsp	
\label{lastpage}
\end{document}